\documentclass{article}

\usepackage{arxiv}

\usepackage[utf8]{inputenc} 
\usepackage[T1]{fontenc}    
\usepackage{hyperref}       
\usepackage{url}            
\usepackage{booktabs}       
\usepackage{amsfonts}       
\usepackage{nicefrac}       
\usepackage{microtype}      
\usepackage{lipsum}
\usepackage{subcaption}
\usepackage{amsmath, amstext, amssymb,bm}
\usepackage{graphicx}

\title{Embedding Hard Physical Constraints in Neural Network Coarse-Graining of 3D Turbulence}

\author{
  Arvind T. Mohan\thanks{Corresponding author: arvindm@lanl.gov} \\
  Center for Nonlinear Studies\\
  Computational Physics and Methods Group \\
  Los Alamos National Laboratory\\
  Los Alamos, NM, United States \\
   \And
 Nicholas Lubbers \\
  Information Sciences Group \\
  Los Alamos National Laboratory\\
  Los Alamos, NM, United States \\
   \And
  Daniel Livescu \\
  Computational Physics and Methods Group \\
  Los Alamos National Laboratory\\
  Los Alamos, NM, United States \\
   \And
  Michael Chertkov \\
  Dept. of Mathematics\\
  University of Arizona\\
  Tucson, AZ, United States \\
}

\begin{document}
\maketitle

\begin{abstract}
In the recent years, deep learning approaches have shown much promise in modeling complex systems in the physical sciences. A major challenge in deep learning of PDEs is enforcing physical constraints and boundary conditions.  In this work, we propose a general framework to directly embed the notion of an incompressible fluid into Convolutional Neural Networks, and apply this to coarse-graining of turbulent flow. These \textit{physics-embedded neural networks} leverage interpretable strategies from numerical methods and computational fluid dynamics to enforce physical laws and boundary conditions by taking advantage the mathematical properties of the underlying equations. We demonstrate results on three-dimensional fully-developed turbulence, showing that this technique drastically improves local conservation of mass, without sacrificing performance according to several other metrics characterizing the fluid flow.
\end{abstract}


\section{Introduction}
\subsection{Deep Learning for High Dimensional Turbulence}

A revolution is underway in computational physics with the promise of deep learning approaches in modeling unresolved physics, accurate coarse-graining and reduced order models (ROMs). This is particularly attractive for fluid flow~\cite{fukami2019super,fukami2019synthetic,kim2020prediction,raissi2019physics,wu2019enforcing,erichson2019physics,wang2019towards,lusch2018deep} and, in particular, turbulence, where the curse of dimensionality coupled with the complexity of the Navier-Stokes equations have hindered much progress over the decades. While recent successes of neural networks (NNs) for turbulence ROMs have gained popularity, we maintain that the curse of dimensionality is as relevant as ever, even in the context of deep learning for physics. There are two key issues with learning high dimensional physics datasets: 1) The computational and memory limitations in employing enough training parameters 2) The black-box nature of NNs that do not guarantee physical laws such as constitutive equations and boundary conditions (BCs). Thus, computational challenges combined with the physics-agnostic nature of NNs can significantly affect our confidence in the learned high-dimensional physics models.

A simple but important example is the continuity equation, which has several applications in physics. For incompressible flows, this equation becomes the divergence-free condition for the velocity field  $V$:
\begin{eqnarray}
\nabla \cdot V \,&=&\, 0 \label{eqn:divV}
\end{eqnarray} 
Consider a standard NN, which is physics-agnostic, trained to model $V$. In the vast space of NN parameter sets, only a small subset of models will produce flows that are consistent with the continuity equation in Eqn.~\ref{eqn:divV}. Some efforts to model turbulence have taken this approach~\cite{fukami2019super,fukami2019synthetic,kim2020prediction}, using deep learning as a black box tool. However, this ignores the continuity equation as a fundamental physical law. A recent approach to incorporate it relies on penalizing the network in the loss function~\cite{raissi2019physics,wu2019enforcing,erichson2019physics,wang2019towards,lusch2018deep} to encourage solutions to obey Eqn.~\ref{eqn:divV} as well as known BC's. We call this approach a \textit{soft constraint} for physics-informed modeling. The loss function encourages the network to find models which are near the physical law, while the overall network remains a black-box. Test examples beyond the scope of training may violate continuity or BC's; as the NN model architecture itself is still blind to the physics. Additionally, the weight of the soft constraint in the loss function becomes an additional hyperparameter that must be tuned during model selection~\cite{bengio2012practical,smith2017cyclical,breuel2015effects}. While soft constraints have been popular due to their flexibility, they provide no guarantees due to lack of an \textit{inductive bias}~\cite{gordon1995evaluation,mitchell1980need,cohen2016inductive,gaier2019weight} in the model; i.e., a model can be expected to perform better if it has built-in knowledge concerning the admissible solution space of a problem. In other words, a model which structurally enforces physical laws is by definition more robust because it offers guarantees on network behavior for data both inside and outside of the training space. Recent work by Wang~\cite{wang2019towards} demonstrated the importance of inductive bias in fluid mechanics machine learning, however, Eqn~\ref{eqn:divV} is still implemented as a soft constraint. While useful, utilizing a soft constraint only approximately realizes these hard physical laws through the difficult-to-interpret non-convex optimization used for training.

Owing to the demonstrated successes of these works for 1D/2D problems, we direct attention to model the full nonlinearity and complexity of turbulence, which is fundamentally a 3D phenomenon~\cite{tennekes1972first} There are significant computational bottlenecks in extending NNs to 3D turbulence datasets for practical applications. This is an acute problem for approaches like PINNs~\cite{raissi2019physics,lu2019deepxde} which need expensive sampling of millions of points for non-chaotic 1D/2D equations. While there has been recent work in adapting NN architectures to learn 3D turbulence, they sacrifice explicit physical constraints~\cite{mohan2019compressed,king2018deep,mohan2018deep,wiewel2019latent} for computational efficiency, raising reliability concerns. A notable work by Kim et. al~\cite{kim2019deep} constrains conservation of mass in 3D fluids, although it does not consider boundary conditions or rigorously evaluate physics of the constraint, since their focus was on computer animation.

\subsection{\label{sec:CNNphysics} Embedding Physical Operators in Convolutional Neural Networks \protect}

Convolutional Neural Networks (CNNs) are well positioned to ingest high dimensional datasets because they utilize \textit{parameter-sharing}~\cite{Goodfellow-et-al-2016}, where a filtering kernel is convolved across the domain to learn the correlation structure in the data. Early layers in the CNN are sensitive to short-range correlations, and later layers build on this to learn long-range correlation as well. Because the same parameters are re-used (shared) in convolutions across the spatial domain, CNNs need orders of magnitude fewer parameters than fully-connected NNs. This is paramount for 2D and 3D flow problems, where the amount of data is exponential in the spatial dimension of the problem, and a fully-connected approach thus requires exponentially more parameters.  The reader is referred to Refs~\cite{lecun2015deep},~\cite{Goodfellow-et-al-2016} and~\cite{krizhevsky2012imagenet} for details of CNNs. Recent physics-informed machine learning has brought into focus the importance of the \textit{structure} of CNN kernels, as explored in PDE-Net~\cite{long2017pde} and related works~\cite{long2019pde,dong2017image,cai2012image}. 

A growing application of CNNs for turbulence is the Convolutional Autoencoder (CAE)~\cite{oja1989neural} for coarse-graining of high fidelity turbulence~\cite{gonzalez2018learning,king2018deep,mohan2019compressed,murata2020nonlinear}. CAEs provide for compression of high-dimensional data like turbulence by using an \textit{encoder} to map the input space into a \textit{latent space} of lower dimensionality by interleaving coarse-graining with learnable convolution layers, and a \textit{decoder} which learns to map this latent space back to the original data in a similar fashion. As this compression forms a natural basis for the construction of ROMs~\cite{mohan2019compressed,hennigh2017lat,erichson2019physics,chen2018collective,wang2019coarse} of turbulent flow and other physical systems, this is a useful domain for research into enforcing hard physical constraints.

The divergence operator $\nabla \cdot$ represents local mass balance, and is a crucial operator in any model of fluid flow. A core aspect of embedding the divergence-free condition as a hard constraint in NN architectures is an accurate and unambiguous definition of an operator which is also amenable to the backpropagation algorithm used during training. Backpropagation through the physics operators ensures that the network is not ``blind" and has intimate knowledge of the constraints through which it must make predictions. To this end, there are three major challenges: \textbf{First}, constructing spatial derivative operators (i.e. $\nabla \times$, $\nabla \cdot$, $\nabla^{2} \cdot$, etc.) that are compatible with the backpropagation algorithm for NN training.  \textbf{Second}, the velocity fields are subject to BCs. \textbf{Third}, the realization of divergence and BC constraints ought to be directly imposed, rather than utilizing soft constraints; We adopt the philosophy of most PDE solvers, where conservation laws and BC constraints are \textit{strictly enforced at all times}. 

In this work, we attempt to address these challenges with a stronger approach which constrains the NN architecture to explore only the physically admissible solution space. We also require that such networks be designed for high-dimensional, fully turbulent flows for practical applicability. The primary motivation is that hard constraints can provide physics guarantees~\cite{ling2016reynolds}, since the NNs are no longer physics-agnostic and the architecture ensures the BCs are prescribed accurately. However, hard constraints are more challenging to implement--the case of embedding  $\nabla \cdot V \,=\, 0$ requires the NN to understand operators such as curl and divergence, while concurrently enforcing BCs. In this work, we report first attempts on a general methodology to address these challenges in a CNN framework with strong inductive bias, as applied to 3D turbulence. Our approach does not modify the number of trainable parameters. Furthermore, it adds explainability by interpreting time-tested strategies from numerical methods and Computational Fluid Dynamics (CFD) as specific instances of CNN kernels, and so can thus be used to design network architectures that include derivative expressions. This offers a way to improve model performance which is completely separate from hyperparameter search. In the next section, we describe our approach for embedding incompressibility into a CNN. In the third section, we describe our experiments and results in implementing this strategy. Finally, we offer conclusions and directions for future work.

\section{\label{sec:CAEarch} Physics Embedded CNN Architecture with Hard Constraints \protect}

\begin{figure*}
    \centering
    \begin{subfigure}[b]{1.0\textwidth}
        \includegraphics[width=\textwidth]{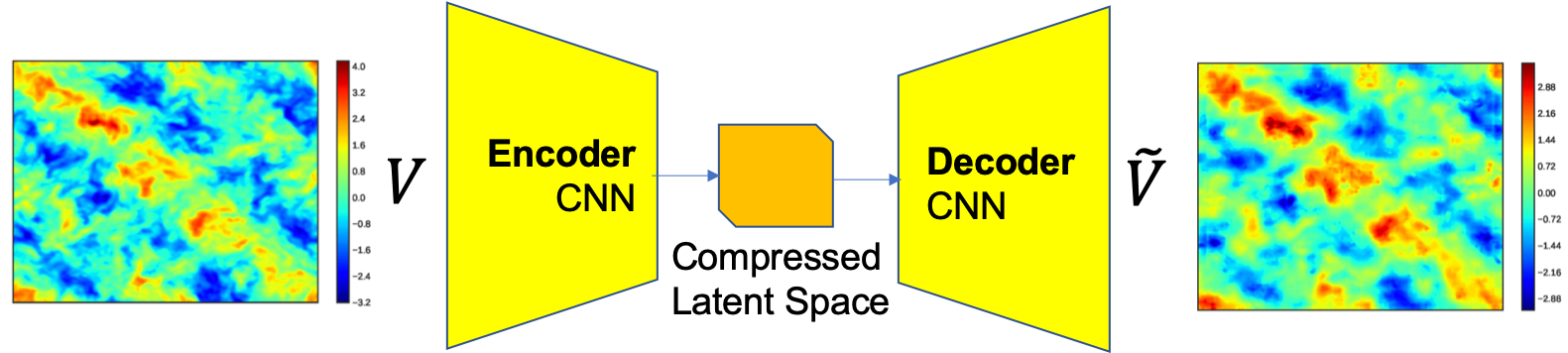}
        \caption{Unconstrained Convolutional Autoencoder with latent space mapping between $V$ and coarse grained $\tilde{V}$ (Slice of U-velocity component for illustration)}
        \label{fig:schematics:CAE}
    \end{subfigure}
    ~ 
      
    \begin{subfigure}[b]{1.0\textwidth}
        \includegraphics[width=\textwidth]{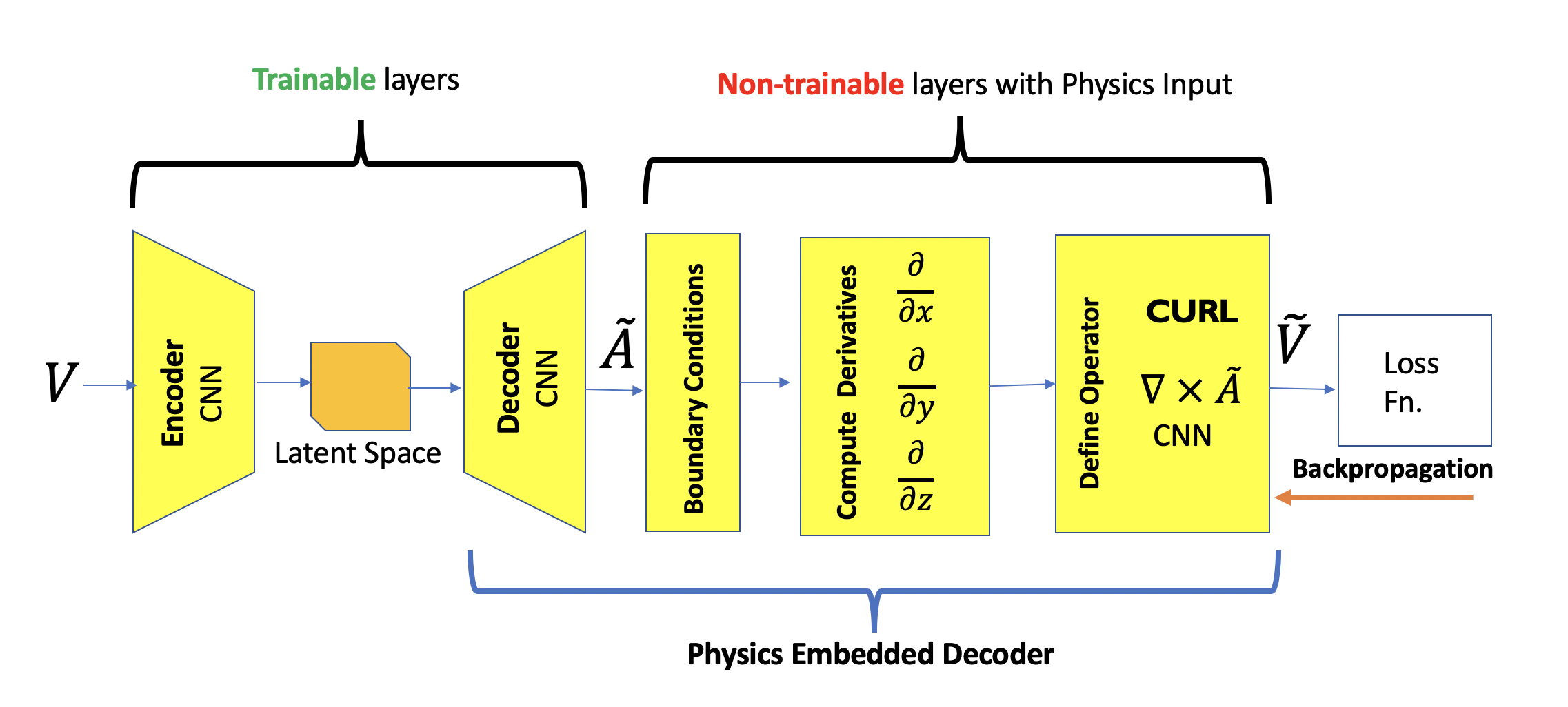}
        \caption{Physics Embedded Convolutional Autoencoder with hard divergence-free constraints for coarse grained $\tilde{V}$}
        \label{fig:schematics:phyCAE}
    \end{subfigure}
\end{figure*}

To demonstrate the aforementioned physics embeddings, we choose a CAE for coarse-graining of high fidelity turbulence. Figure~\ref{fig:schematics:CAE} shows the standard, physics-agnostic CAE architecture, with the 3D velocity field training dataset as input $V$. The CAE encoder learns a compressed latent space, significantly smaller than $V$. The velocity field is reconstructed from the latent space by a CAE decoder that is learned simultaneously with the encoder. However, this data compression is accompanied by some loss of information, and therefore the reconstructed $V$ is denoted as $\tilde{V}$. Therefore we can consider $\tilde{V}$ as a coarse-grained approximation of $V$. Since the CAE is unconstrained, the challenge is to embed physical constraints to ensure that the CAE only learns a latent space such that $\nabla \cdot \tilde{V} \,=\, 0$.

Since embedding Eqn.~\ref{eqn:divV} requires computing derivatives on a field that resides on a discretized solution mesh, it is pragmatic to adopt strategies from well-known finite difference (FD)/Finite volume (FV) numerical methods. These FD/FV algebraic discretizations can be analytically derived from Taylor series expansions around a fixed point~\cite{ferziger1981numerical, spalding1972novel}. These discretizations can further be expressed in matrix form called \textit{stencils}. Upon examination, it is apparent that the CNN kernels are \textit{structurally equivalent} to the FV stencils~\cite{long2017pde,dong2017image,long2019pde}. This simple, but powerful connection allows us to embed these stencils as CNN layers with fixed kernel weights to compute spatial derivatives. For details on this approach, see  Appendix~\ref{app:CNNphysics:customKernel}. We employ a technique to enforce BCs in the CNN using the idea of \textit{ghost cells}~\cite{fadlun2000combined,tseng2003ghost,berthelsen2008local} from CFD, as outlined in Appendix~\ref{app:CNNphysics:BC}. This allows us to implement BCs to a given order of discretization accuracy. Both these techniques involve a close fusion of concepts from neural networks, numerical methods and CFD.

Having implemented spatial derivatives, we model the flow using a potential formulation~\cite{hirasaki1968general,morino1985scalar,biro1989use} based on the Helmholtz decomposition with vector potential $A$ and scalar potential $\psi$ that models the flow:

\begin{eqnarray}
V \,&=&\, \nabla\times A+\nabla \psi 
\end{eqnarray}

The divergence constraint annihilates all terms associated with the vector potential $A$, giving rise to a Laplace constraint only on the scalar potential: $\nabla^2\psi=0$.  For steady boundary conditions, the scalar potential corresponds to a steady background flow, which can be subtracted from the data so that the learning problem is restricted to the turbulent fluctuations. In particular, for data analyzed here, the boundary conditions are periodic, so $\psi=0$ is a valid solution.
The key idea is as follows: Instead of only predicting a velocity field, we choose to make an intermediate prediction for vector potential $\tilde{A}$, while framing the final network prediction $\tilde{V}$ in the target velocity space via the curl, implemented as a  numerical stencil:
\begin{eqnarray}
V \,&=&\, \nabla \times A \label{eqn:curlA} 
\end{eqnarray}

Then, predictions $\tilde{V}$ will \textit{automatically} obey Eqn.~\ref{eqn:divV} up to the accuracy of the stencil. Figure~\ref{fig:schematics:phyCAE} shows the autoencoder with a  \textbf{physics-embedded CNN AutoEncoder} (PhyCAE) where this strategy is implemented, in addition to those in Appendix~\ref{app:CNNphysics:customKernel}. The CAE encoder learns a latent space from $V$, from which the decoder constructs $\tilde{A}$.
While the flow dataset itself does not directly contain $A$, (and $A$ itself can only be defined up to gauge transformations), the network can implicitly learn $\tilde{A}$; we constrain the learning by requiring that the curl of the decoder prediction $\tilde{A}$ be equal to $\tilde{V}$. The physics is thus embedded in the modified decoder, which consists of a CNN layer enforcing BCs with ghost cells (Appendix~\ref{app:CNNphysics:BC}), followed by another CNN layer which computes all the spatial derivatives necessary to define a curl operator (Appendix~\ref{app:CNNphysics:customKernel}). The last CNN layer uses these to compute the curl operator on the $\tilde{A}$ field. Therefore, all layers after the decoder CNN in the PhyCAE are non-trainable and fully explainable, as they merely construct Eqn.~\ref{eqn:curlA} with numerical methods.

%

\section{\label{sec:results} Results \protect}

\subsection{Training \label{sec:results:training}}
To illustrate the performance of the physics embedded autoencoders, we train two cases: The standard CAE with zero padding in Fig.~\ref{fig:schematics:CAE}, and the PhyCAE in Fig.~\ref{fig:schematics:phyCAE}.  The dataset is high-fidelity direct numerical simulation (DNS) of a 3D Homogeneous Isotropic turbulence (HIT), which is described in Appendix~\ref{app:dataset} and Ref.~\cite{daniel2018reaction}.

The CAE architecture has 3 layer encoder-decoder with an ADAM optimizer and L2 loss, with $6$ filters at each level, and was evaluated for turbulence in Mohan~\cite{mohan2019compressed}. We intentionally use fewer filters to avoid over-training and illustrate the effects of physics input in the NNs. In this architecture,  the compression ratio, defined as the ratio between the dimension of a single datapoint ($3 \times 128^{3}$) and that of the latent space ($6 \times 15^{3}$), is $\approx 300$.  The PhyCAE architecture comprises of the CAE layers above with the same hyperparameters, but with the addition of the physics embedded layers in the decoder. We emphasize again that these layers add no extra learnable parameters to the network. Since the dataset is statistically stationary with a total of 12 eddy turnover times, we train both networks on 0 - 0.75 eddy times. The test dataset is from 4 - 4.75 eddy times, so that we test if the network has learned the statistical behavior of the fluid correlations at future time. We also note that all calculations here are performed with single precision, hence the relative numerical precision is $\approx 10^{-7}$.

\subsection{Effect of Hard Constraints on Training \label{sec:results:constraints}}
In this section, we investigate how stringently CAE and PhyCAE adhere to the $ \nabla \cdot V$ constraint, and explore the characteristics of PhyCAE from the perspective of accuracy, computational efficiency, training speed and reliability.

\begin{figure}[h]
\centering
\includegraphics[scale=0.25]{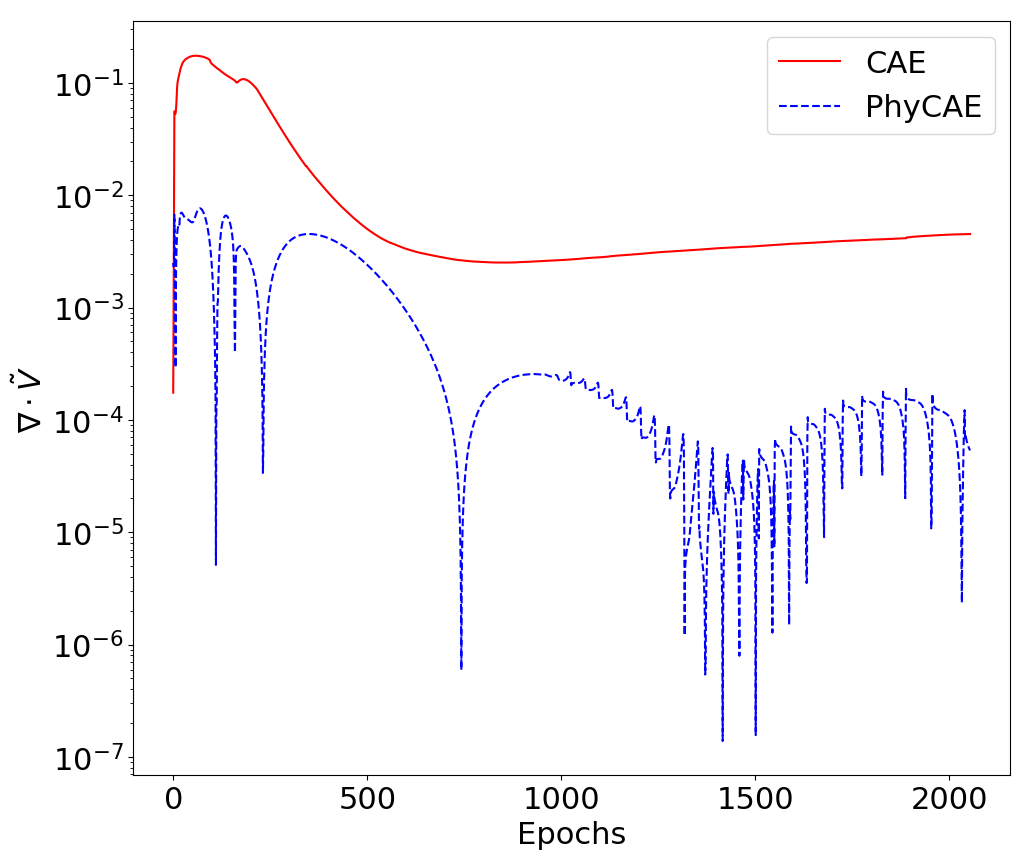}
\caption{\label{fig:results:divergenceCompare} Variation of $\nabla \cdot \tilde{V}$ with training epochs for CAE vs. PhyCAE. \textbf{Final Divergence:} $\mathbf{ \approx  10^{-5}}$ for PhyCAE vs $\mathbf{\approx  10^{-2}}$ for CAE}
\end{figure}

NNs make predictions every epoch, and if the training loss converges to a minima with (typically several thousand) more epochs, we expect the predictions to have ``learned" some physical laws. One of the key expectations from any physics-embedded hard constraint, is that the network must be cognizant of these physical laws \textit{from the very first epoch} and the predictions never deviate far from the imposed constraints i.e. the stay on the physical subspace. In other words, while improved optimization methods (adaptive regularization, optimizers, optimal hyperparameters, minimization tricks etc.) can assist the physics embeddings in finding minima with the desired physical laws, the hard constraints should ensure that the network predictions stay in the physically constrained subspace \textit{by design} and not solely by learning. As a result, we should expect PhyCAE to find minima faster and be more consistent with the physics than the CAE. To quantify how well the constraint is realized, we measure the total absolute divergence (TAD) across each example averaged over the examples, given by the formula $\sum |\nabla \cdot \tilde{V}|$. A model which perfectly enforces incompressible flow minimizes this quantity. However, the TAD is not zero even in the PhyCAE due to the second order discretization stencils used to apply derivative operators and the limitations of single-precision floating-point arithmetic. 

Figure~\ref{fig:results:divergenceCompare} shows the TAD on the training data as a function of the network training time for both the CAE and PhyCAE. For CAE, we see a spike in the divergence as high as $\sim 10^{-1}$, and approaches to a final value of $\sim 10^{-2}$, although it appears that it might increase with additional training. The TAD for the PhyCAE starts at $\sim 10^{-2}$ and trends downward during training, even oscillating near numerical zero, and typically between $10^{-4}$ and $10^{-5}$. In other words, the \textit{best-case} for the CAE is comparable to the \textit{worst-case} for the PhyCAE, and after training, the PhyCAE performs more than 2 orders of magnitude better with respect to TAD.  We have likewise computed the TAD after training on the test dataset after training. In this case, the TAD for the PhyCAE is $\sim 10^{-5}$, while CAE is $\sim 10^{-2}$, further emphasizing the generality and robustness of our physics embeddings. The physics-aware inductive bias of the PhyCAE allows it to perform far better than the CAE while training with the identical hyperparameters, and number of trainable parameters.
We remark that much like any PDE solver, the discretization errors affect the accuracy of the hard constraint, such as the 2\textsuperscript{nd} order scheme used in this work. Due to the interpretable hard-constraint approach of the PhyCAE, this could be further decreased by improving the spatial discretization method in Eqn.~\ref{eqn:centralDiff} from a 2\textsuperscript{nd} to a higher order scheme. This extension is straightforward since the CNN allows for kernels of larger sizes produced by higher order numerical schemes. This would require a corresponding change in number of ghost cells, which can be implemented as outlined in Eqn.~\ref{eqn:MoreGhostCells} in Appendix~\ref{app:CNNphysics:BC}.

\subsection{Turbulence Diagnostics \label{sec:results:diagnostics}}
We now briefly describe 3 important tests of 3D turbulence which are used as diagnostic metrics for the accuracy of the flow predicted by the trained model. While these metrics do not explicitly address the divergence of the velocity, they target specific properties of turbulence and are important to have confidence in the accuracy of the predicted model. 

\subsubsection{$4/5$ Kolmogorov law and the Energy Spectra}

The main statement of the Kolmogorov theory of turbulence is that asymptotically in the inertial range, i.e. at $L\gg r\gg\eta$, where $L$ is the largest, so-called energy-containing scale of turbulence and $\eta$ is the smallest scale of turbulence, so-called Kolmogorov (viscous) scale, $F(r)$ does not depend on $r$.  Moreover, the so-called $4/5$-law states for the third-order moment of the longitudinal velocity increment
\begin{eqnarray}
&& L\gg r\gg\eta:\quad S_3^{(i,j,k)}\frac{r^i r^j r^k}{r^3}=-\frac{4}{5}\varepsilon r,
\label{eq:4/5}
\end{eqnarray}
where $\varepsilon=\nu D_2^{(i,j;i,j)}/2$ is the kinetic energy dissipation also equal to the energy flux. Self-similarity hypothesis extended from the third moment to the second moment results in the expectation that within the inertial range, $L\gg r\eta$, the second moment of velocity increment scales as, $S_2(r)\sim v_L (r/L)^{2/3}$. This feature is typically tested by plotting the energy spectra of turbulence (expressed via $S_2(r)$) in the wave vector domain. 

\begin{figure}[h]
\centering
\includegraphics[scale=0.4]{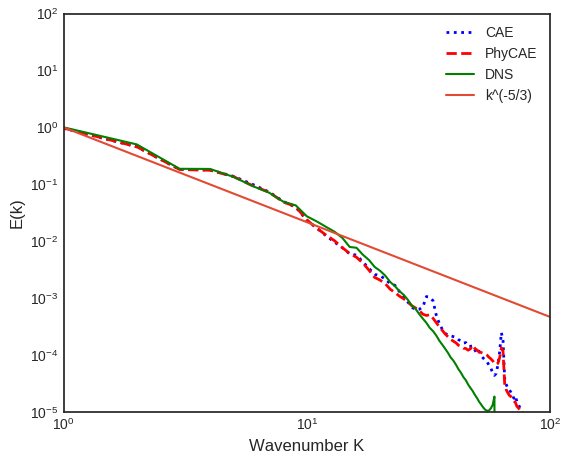}
\caption{\label{fig:results:energySpec} Kolmogorov Energy Spectra of $\tilde{V}$ for  CAE and PhyCAE vs DNS $V$ }
\end{figure}

Figure~\ref{fig:results:energySpec} shows the energy spectra of the coarse grained $\tilde{V}$  predicted by the CAE and PhyCAE  compared with the ground truth value from the DNS test data $V$. The results show excellent large scale (low wavenumbers)  and inertial range accuracy by the PhyCAE, very similar to that of CAE. Most of the discrepancies are localized at the small scales (high wavenumbers), due to the information loss that occurs during coarse graining.  However, since most practical applications of ROMs focus only on large/inertial scales, we find these discrepancies acceptable.

\subsubsection{Intermittency of Velocity Gradient}

Consequently from Eqn.~\ref{eq:4/5}, the estimation of the moments of the velocity gradient results in
\begin{eqnarray}
&& D_n\sim \frac{S_n(\eta)}{\eta^n}.
\label{eq:D_n_scaling}
\end{eqnarray}
This relation is strongly affected by intermittency for large values of $n$ (i.e. extreme non-Gaussian behavior) of turbulence, and is a valuable test of small scale behavior.

\begin{figure}[h]
\centering
\includegraphics[scale=0.4]{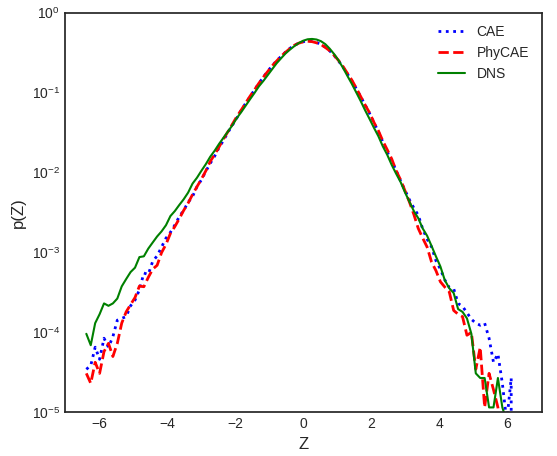}
\caption{\label{fig:results:velocitypdf} PDFs of velocity gradient of $\tilde{V}$ for  CAE and PhyCAE vs DNS $V$}
\end{figure}
The velocity gradient PDFs thus described are computed for both $\tilde{V}$ and $V$ , shown in Fig.~\ref{fig:results:velocitypdf}. Once again, we see excellent matches between predicted PhyCAE and DNS velocities, with minor discrepancies in the tails, which are likely a manifestation of the small scale errors. CAE is quite close to PhyCAE in observed accuracy, with the latter being slightly more accurate in the tails.

\subsubsection{Statistics of coarse-grained velocity gradients: $Q-R$ plane.}

Both the diagnostics described so far are highly averaged quantities which do not explicitly account for the flow structure and 3D effects of the predicted velocities. To address this, we utilize isolines of probability in the $Q-R$ plane, which expresses intimate features of the turbulent flow topology, having a nontrivial shape documented in literature. See Ref.~\cite{chertkov1999lagrangian} and references therein. Different parts of the $Q-R$ plane are associated with different structures of the flow. Thus, lower right corner (negative $Q$ and $R$), which has higher probability than other quadrants, corresponds to a pancake type of structure (two expanding directions, one contracting) with the direction of rotation (vorticity) aligned with the second eigenvector of the stress. This tear-drop shape of the probability isoline becomes more prominent with decrease of the observation scale $r$. Here, we study the $Q-R$ plane filtered at different scales $r$, to examine large ($r=32$), inertial ($r=8$), and small scale ($r=1$) behaviors. This allows us to selectively analyze the accuracy of our predictions at different scales, since we are interested in modeling primarily the large and inertial ranges.

\begin{figure}[h]
\centering
\includegraphics[scale=0.3]{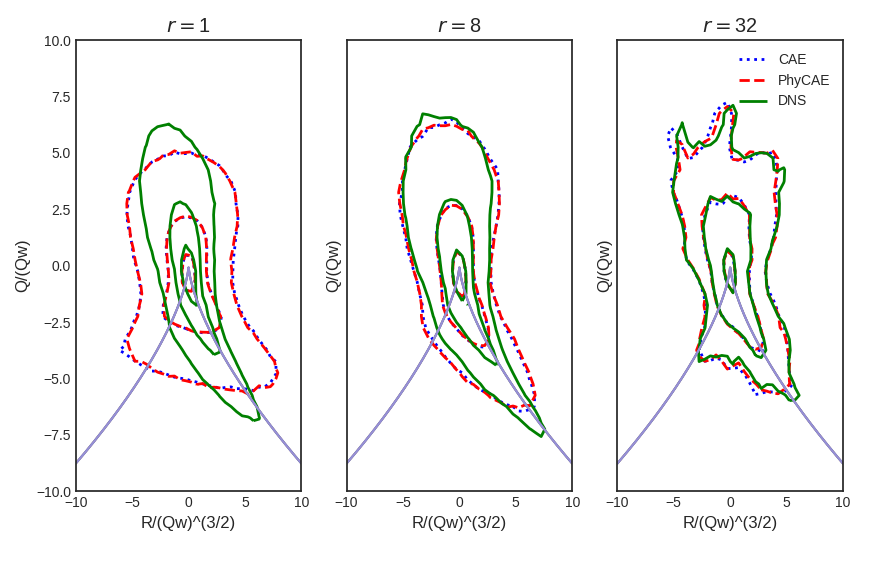}
\caption{\label{fig:results:qr} $Q-R$ plane PDFs of $\tilde{V}$ for  CAE and PhyCAE vs DNS $V$}
\end{figure}

The $Q-R$ plane PDFs in Fig.~\ref{fig:results:qr} show a clear picture of the PhyCAE predictive capability.  Both the networks show that large scale flow topology is captured extremely well, with inertial scales having minor - but acceptable - discrepancies. The small scale error is quite significant, as seen from the Kolmogorov spectra and velocity PDF. Overall, the $Q-R$ plane shows that we are indeed capable of predicting coarse grained 3D fields that have excellent large scale resolution. 

We note that the unconstrained CAE also produces results of comparable accuracy, but lacks in conserving continuity. On the other hand, PhyCAE shows excellent predictive capability while also closely adhering to the divergence-free criterion as outlined in the previous section.

\section{\label{sec:conclusion} Conclusion \protect}

Our work introduces and demonstrates the effectiveness of a simple and interpretable method of enforcing incompressibility of a fluid flow within CNNs. Since this is enforced as a structural aspect of the CNN, the method is a hard constraint rather than a soft constraint, and so yields no additional hyperparameters to tune. Another useful consequence is that the constraint is enforced through the training procedure.  This is accomplished by utilizing static, non-trainable layers for spatial derivatives, whose kernels are stencils that can be derived using the standard discretizations. Likewise, we implement boundary conditions by modifying the padding operation of a CNN. Using these tools, train the decoder to output a vector potential that describes perturbations around a background flow. We demonstrate this scheme for a CAE architecture; however, it is generally applicable to any task where a CNN should output an incompressible velocity field, for example, flow forecasting~\cite{mohan2019compressed,hennigh2017lat} or generative modeling of flows. While this work is directly applicable to uniform and structured meshes, a useful avenue of research would be extension to non-uniform and unstructured meshes, possibly exploiting recent developments in GraphCNNs~\cite{wang2018dynamic}.

An interesting feature of this scheme is the gauge ambiguity of the vector potential $\tilde{A}$. In traditional vector potential approaches to PDEs, a gauge invariance needs to be resolved by fixing a gauge in which computations take place. Our approach did not specify a gauge, and future work might explore the effective gauge conditions learned by the model and characterize them in relation to the classical body of work on gauge fixing. Problems involving incompressible fluids are common, but further work could seek to extend the context of this strategy. Thus, future work could focus on addressing more general constraints of the form
\begin{eqnarray}
L(V) = 0,
\end{eqnarray}
for more general differential operators $L$ and physical fields $V$, by enforcing them as CNN layers; work on soft constraints has identified this as a general target for physics-informed machine learning~\cite{raissi2019physics,lu2019deepxde}, and we concur with this direction. Research addressing such problems with hard constraints (i.e. through structural aspects of model architecture) would make machine learned models of a great many physical systems more robust and interpretable.

\section{Acknowledgements}
The authors thank Don Daniel for generating the DNS data and also thank Jonah Miller, Vitaliy Gyrya, Peetak Mitra and Srinivasan Arunajatesan for useful discussions. This work has been authored by employees of Triad National Security, LLC which operates Los Alamos National Laboratory (LANL) under Contract No. 89233218CNA000001 with the U.S. Department of Energy/National Nuclear Security Administration. A.T.M. and D.L. have been supported by LANL's LDRD program, project number 20190058DR. A.T.M also thanks the Center for Nonlinear Studies at LANL for support and acknowledges the ASC/LANL Darwin cluster for GPU computing infrastructure.

\section*{Appendix}
\label{appendix}
\appendix

\section{\label{app:dataset} 3D Homogeneous Isotropic Turbulence Dataset \protect}

\begin{figure}[h]
    \centering
    \begin{subfigure}[b]{0.3\textwidth}
        \includegraphics[width=\textwidth]{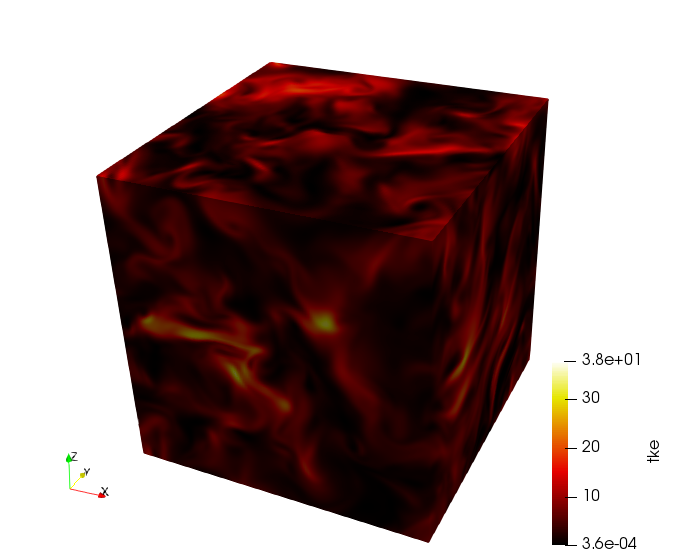}
        \caption{Instantaneous turbulent kinetic energy}
        \label{fig:TKE}
    \end{subfigure}
    ~ 
    \begin{subfigure}[b]{0.3\textwidth}
        \includegraphics[width=\textwidth]{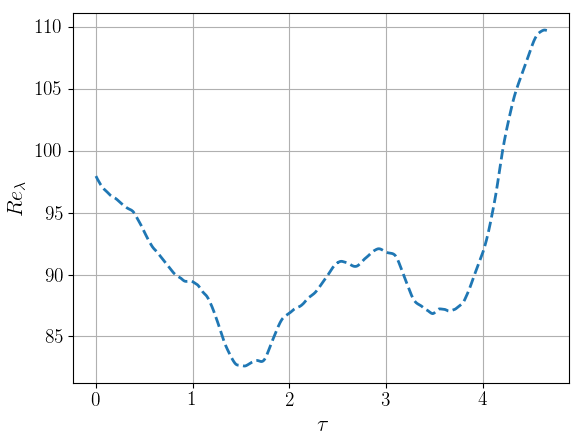}
        \caption{Reynolds number (based on Taylor microscale)}
        \label{fig:Re}
    \end{subfigure}
    ~ 
    \begin{subfigure}[b]{0.3\textwidth}
        \includegraphics[width=\textwidth]{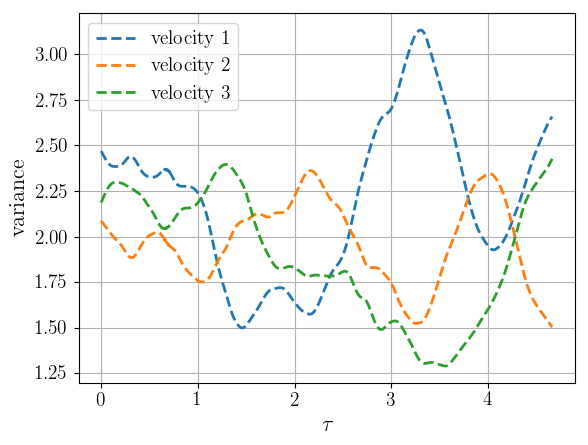}
        \caption{Individual velocity variances}
        \label{fig:velocityVar}
    \end{subfigure}
    \caption{Representative Statistics of the Simulation}\label{fig:dataset}
\end{figure}

The dataset consists of a 3D Direct Numerical Simulation (DNS) of homogeneous, isotropic turbulence, in a box of size $128^{3}$. We denote this dataset as \textit{HIT} for the remainder of this work.  We provide a brief overview of the simulation and its physics in this section, and a detailed discussion can be found in Daniel et. al~\cite{daniel2018reaction}.
The ScalarHIT dataset is obtained using the incompressible version of the CFDNS~\cite{livescu2009cfdns} code, which uses a classical pseudo-spectral algorithm. We solve the incompressible Navier-Stokes equations:

\[
\partial_{x_{i}}v_{i}=0,\qquad\partial_{t}v_{i}+v_{j}\partial_{x_{j}}v_{i}=-\frac{{1}}{\rho}\partial_{x_{i}}p+\nu\Delta v_{i}+f_{i}^{v},
\]
where $f^{v}$ is a low band forcing, restricted to small wavenumbers
$k<1.5$ {[}1{]}. The $128^{3}$ pseudo-spectral simulations are dealiased
using a combination of phase-shifting and truncation to achieve a
maximum resolved wavenumber of $k_{max}=\sqrt{{2}}/3\times128\sim60$.

For illustration, Figure~\ref{fig:TKE} shows the turbulent kinetic energy at a time instant. Figure~\ref{fig:Re} shows the variation in the Taylor-microscale based Reynolds number with the eddy turnover time, which characterizes the large turbulence scales. Finally, the variances in all $3$ velocity components are shown in Fig.~\ref{fig:velocityVar}. Based on the sampling rate, each eddy turnover time $\tau$ consists of 33 snapshots. The training dataset uses $22$ snapshots $\approx 0 -0.75 \tau$ and test dataset also consists of 22 snapshots in $\approx 4 -4.75 \tau$.

\section{Spatial Derivative Computation in CNN Kernels \label{app:CNNphysics:customKernel}}

\begin{figure}[h]
\centering
\includegraphics[scale=0.5]{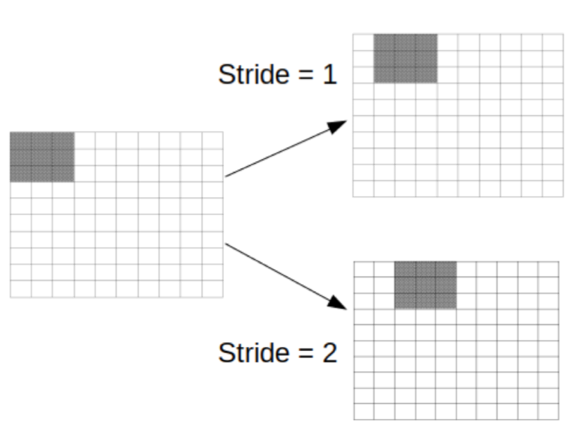}
\caption{\label{fig:schematics:striding} Illustration of 1-step and 2-step striding mechanisms}
\end{figure}

In deep learning, CNNs are used to learn the spatial features with a (symmetric or non-symmetric) kernel $f$ as a convolution operation throughout the domain of interest $g$ at a layer $n$. The $n^{th}$ CNN layer thus computes a field $y$ as $ y_{n} \,=\, f * g_{n-1}$, where $g_{n-1}$ is the domain computed by the $(n-1)^{th}$ layer. Therefore, $y_{n} \,=\, g_{n}$  at layer $n+1$, such that $ y_{n+1} \,=\, f * g_{n}$. The kernel translation is also an important hyperparameter, called \textit{striding}, that can be performed for every point in the mesh (1-step), or by skipping over a two or more points every time the kernel is translated (2-step and higher), as shown in the illustration in Fig.~\ref{fig:schematics:striding}. Subsequently, these kernel weights are iteratively learned by backpropagation and gradient-based optimization.

Since we intend to compute derivatives on a field that resides on a discretized solution mesh, it is pragmatic to adopt strategies from well-known finite difference (FD)/Finite volume (FV) numerical methods to compute derivatives, which can be analytically derived from Taylor series expansions around a fixed point~\cite{ferziger1981numerical, spalding1972novel}. These forward, backward, upwind or central difference discrete numerical approximations of a continuous derivative can be approximated to a desired $m$\textsuperscript{th} order~\cite{morinishi1998fully, visbal2002use} of accuracy.  Consider a variable of interest $\phi$ that resides on a $3D$ mesh.  A standard 2\textsuperscript{nd} order central difference FV scheme for partial derivatives is shown in Eqn.~\ref{eqn:centralDiff}. Its coefficients can be expressed as a  matrix, called a numerical \textit{stencil} as shown in Eqn.~\ref{eqn:stencil}. 

\begin{eqnarray}
\frac{\partial \phi}{\partial x} &= \frac{\phi(x + \delta x) - \phi (x - \delta x)}{2 \delta x} + O(\delta x)^{2} \nonumber \\
\frac{\partial \phi}{\partial y} &= \frac{\phi(y + \delta y) - \phi (y - \delta y)}{2 \delta y} + O(\delta y)^{2} \nonumber \\
\frac{\partial \phi}{\partial z} &= \frac{\phi(z + \delta z) - \phi (z - \delta z)}{2 \delta z} + O(\delta z)^{2} 
\label{eqn:centralDiff}
\end{eqnarray}

\begin{eqnarray}
 \frac{\partial \phi}{\partial x} &=
  \begin{bmatrix}
    0 & 0 & 0\\[0.5em]
    - \frac{1}{2 \delta x} & 0 & \frac{1}{2 \delta x}\\[0.5em]
    0 & 0 & 0
  \end{bmatrix}   \nonumber \\
  \frac{\partial \phi}{\partial y} &=
  \begin{bmatrix}
   0& - \frac{1}{2 \delta y} & 0\\[0.5em]
    0 & 0 & 0\\[0.5em]
    0& \frac{1}{2 \delta y} & 0 
  \end{bmatrix}  \nonumber \\
  \frac{\partial \phi}{\partial z} &=
  \begin{bmatrix}
    0 & 0 & 0\\[0.5em]
    - \frac{1}{2 \delta z} & 0 & \frac{1}{2 \delta z}\\[0.5em]
    0 & 0 & 0
  \end{bmatrix}
\label{eqn:stencil}
\end{eqnarray}

Upon examination, it is apparent that the CNN kernels are \textit{structurally equivalent} to the FV stencils~\cite{long2017pde,dong2017image,long2019pde}. Therefore, FV stencils are essentially CNN kernels with fixed, non-trainable weights that compute a derivative to the desired order of accuracy. Furthermore, the CNN kernel and numerical stencil operations are \textit{mathematically identical} in the context of 1-step striding. Similar to the CNN convolution operator above, domain $g$ is the numerical mesh and $f$ is the FV kernel. The derivative $y^{\prime}$ is then computed as $y^{\prime} \,=\, f * g$, at any layer $n$. Subsequent layers can be constructed to compute higher order derivatives with appropriate stencil kernels.

This simple, but powerful, connection allows us to embed these stencils as CNN layers with fixed kernel weights to compute our derivatives of interest. Furthermore, we can explicitly define the order of accuracy in derivative computation by leveraging higher order numerical schemes. This allows us to trade-off compute costs and accuracy while simultaneously being \textit{interpretable}.

\section{Enforcing Periodic Boundary Conditions \label{app:CNNphysics:BC}}

Boundary conditions are a critical physics component that any computational approach - traditional PDE solvers or Deep Learning based - must rigorously impose in order to preserve the physics of the flow. The HIT flow has spatially periodic BCs in all three directions. In order to perform realistic physics embedding in CNNs, we present here a method to enforce BCs in CNNs, to a desired order of discretization accuracy.

\begin{figure}[h]
\centering
\includegraphics[scale=0.5]{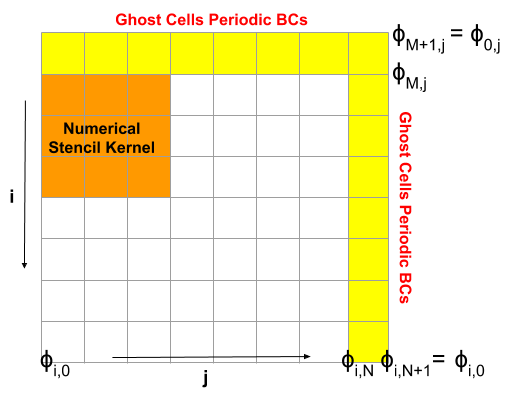}
\caption{\label{fig:schematics:BCpadding1} CNN kernel computing numerical derivative in interior nodes (also shown periodic BCs prescribed as ghost cells)}
\end{figure}

\begin{figure}[h]
\centering
\includegraphics[scale=0.5]{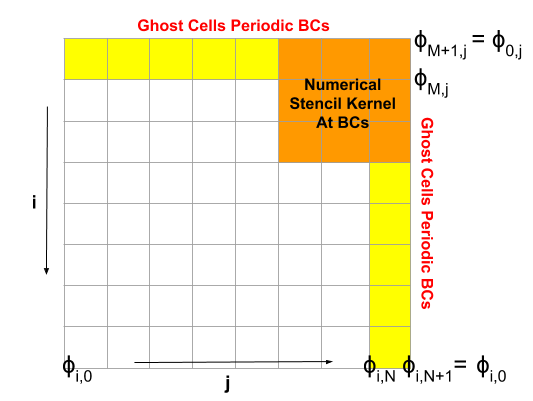}
\caption{\label{fig:schematics:BCpadding2} CNN kernel computing numerical derivative at ghost cells boundary nodes with periodic BCs}
\end{figure}

Figure~\ref{fig:schematics:BCpadding1} shows the CNN kernel defined by a numerical stencil as described in Appendix~\ref{app:CNNphysics:customKernel}. The $(3 \times 3 \times 3)$ kernel performs convolution on the mesh, with 1-step striding. As a result, the outermost column/row of cells in the domain are forfeited. Due to the loss of boundary cells, the kernel is not able to accurately compute the derivative with the BCs, which can cause significant inaccuracies in the solution.  In the machine learning community, a popular fix is to recover the original dimensionality by ``padding" these dimensions by a constant valued scalar (typically zero) or replicate with the value of adjacent cells. However, this is a well-known issue in the CFD community, and is more pronounced when using higher order numerical stencils~\cite{fadlun2000combined} with kernel sizes of $(5 \times 5 \times 5)$ or $(7 \times 7 \times 7)$. A standard approach to resolve this discrepancy while simultaneously satisfying the BCs, is by employing \textit{ghost cells}~\cite{fadlun2000combined,tseng2003ghost,berthelsen2008local}. Ghost cells are ``virtual" cells which are defined at mesh boundaries, so that a derivative of the desired order consistent with the numerical stencil can be computed. In the solution domain illustrated in Fig.~\ref{fig:schematics:BCpadding1},  $x$, $y$ and $z$ have dimensions of $N+1$, $M+1$ and $L+1$ respectively, which reduce to $N$, $M$ and $L$ due to the $(3 \times 3 \times 3)$ convolution kernel. Therefore, $i^{th}$ direction now ranges from $0 \rightarrow M$, the $j^{th}$ from $0 \rightarrow N$ and $k^{th}$ direction from $0 \rightarrow L$. The $k^{th}$ direction in the 3D domain is not shown here for convenience. Periodic BCs imply the flow leaving the domain in one direction enter the domain in the opposite direction. Ghost cells $(N+1)$ are now created to mimic this behavior at the boundaries , shown as the yellow cells in Fig.~\ref{fig:schematics:BCpadding1}. The solution at $(N+1)$ is set as $\phi_{i,N+1,k} \,=\, \phi_{i,0,k}$ which exactly satisfies the periodic BC constraint. Likewise in the $i^{th}$ axis, $\phi_{M+1,j,k} \,=\, \phi_{0,j,k}$ and $\phi_{i,j,L+1} \,=\, \phi_{i,j,0}$ are the BCs in the $k^{th}$ axis. 

Figure~\ref{fig:schematics:BCpadding2} shows the net effect of the ghost cells, with the CNN kernel computing derivatives with BCs. It is important to note that the ghost cells are flexible for higher order numerical schemes as well, by adding more cells to account for the larger kernel. For instance, in the $j^{th}$ axis we can add 3 columns of ghost cells with periodic BCs such that,

\begin{eqnarray}
\phi_{i,N+1,k} \,=\, \phi_{i,0,k} \nonumber \\
\phi_{i,N+2,k} \,=\, \phi_{i,1,k} \nonumber \\
\phi_{i,N+3,k} \,=\, \phi_{i,2,k} 
\label{eqn:MoreGhostCells}
\end{eqnarray}

We can thus specify BCs via ghost cells by developing a custom padding based on the desired order of numerical accuracy.

\bibliographystyle{unsrt}  


\end{document}